\documentstyle[12pt]{article}
\title{\bf A Correction to the Hamiltonian of the QCD String with Quarks
 due to the
Rigidity Term}
\author{D.V.ANTONOV \thanks{E-mail:
antonov@pha2.physik.hu-berlin.de; on leave of
absence from the Institute of Theoretical and Experimental Physics (ITEP);
supported by Graduiertenkolleg {\it Elementarteilchenphysik}, 
Russian Fundamental Research Foundation, Grant 
No.96-02-19184, DFG-RFFI, Grant 436 RUS 113/309/0  
and by the INTAS, Grant No.94-2851.}
\\
{\it Institut f\"ur Elementarteilchenphysik, Humboldt-Universit\"at,}\\
{\it Invalidenstrasse 110, D-10115, Berlin, Germany}\\}  
\date{}
\begin{document}
\maketitle
\vspace{1cm}
\centerline{\bf {Dedicated to the memory of Professor M.V.Terentiev}}
\vspace{1cm}
\centerline{\bf {Abstract}}
\vspace{3mm}
The action of the gluodynamics string, 
obtained in Ref.1, is used for the   
derivation of the correction, arising due to the rigidity term, to the 
Hamiltonian of the quark-antiquark system, which was 
obtained in Refs.2 and 3.   
This correction contains additional contributions to
the orbital momentum of the system
and several higher derivative operators. With the help of the obtained 
Hamiltonian 
a rigid string-induced term in the Hamiltonian of the relativistic quark
model is evaluated for the case of large masses of a quark and 
antiquark.
\newpage
{\large \bf 1. Introduction}

\vspace{3mm}
In a recent paper$^{1}$ it was shown that the effective action of the 
gluodynamics string obtained from the expansion of the averaged Wilson
loop $\left<W(C)\right>$, written through the non-Abelian Stokes 
theorem$^{4,5}$ and
cumulant expansion$^{5,6}$, has the form of a series in powers of 
$\frac{T_g}{L}$, where $T_g$ is the correlation length of the vacuum,
and $L$ is the size of the Wilson loop. Keeping in the cumulant 
expansion only
the lowest-- bilocal term, which is dominant according to lattice
data$^{7}$,

$$\left<W(C)\right>=tr~\exp \left(-\int\limits_S^{}d\sigma_{\mu\nu}(w)
\int\limits_S^{}d\sigma_{\lambda\rho}(w^\prime)
\left<F_{\mu\nu}(w)\Phi(w,w^\prime)F_{\lambda\rho}(w^\prime)\Phi(w^\prime,w)
\right>
\right),$$
and parametrizing it in the following way$^{7,8}$

$$\left<F_{\mu\nu}(w)\Phi(w,w^\prime)F_{\lambda\rho}(w^\prime)
\Phi(w^\prime,w)\right>=
\frac{\hat{1}}{N_c} \Biggl\{\left(\delta_{\mu\lambda}\delta_{\nu\rho}-
\delta_{\mu\rho}\delta_{\nu\lambda}\right)D\Biggl(\frac{(w-w^\prime)^2}{T_g^2}
\Biggr)+$$

$$ +\frac{1}{2}\Biggl[\frac{\partial}{\partial w_\mu}\left(
(w-w^\prime)_\lambda
\delta_{\nu\rho}-(w-w^\prime)_\rho\delta_{\nu\lambda}\right)+\frac{\partial}
{\partial w_\nu}\left(
(w-w^\prime)_\rho\delta_{\mu\lambda}-(w-w^\prime)_\lambda
\delta_{\mu\rho}\right)
\Biggr]D_1\Biggl(\frac{(w-w^\prime)^2}{T_g^2}\Biggr)\Biggr\},
$$
where $D$ and $D_1$ are two renormalization group invariant coefficient 
functions, one arrives at the following effective action of the gluodynamics
string, induced by the nonperturbative background fields

$$S_{biloc.}=-\ln\left<W(C)\right>=\sigma\int d^2\xi\sqrt{g}+\frac{1}{\alpha_0}
\int d^2\xi\sqrt{g}g^{ab}\left(\partial_at_{\mu\nu}\right)\left(
\partial_bt_{\mu\nu}\right)+
O\Biggl(\frac{T_g^6}{L^2}\alpha_s tr\left<F_{\mu\nu}^2(0)\right>\Biggr)
, \eqno(1)$$
where

$$\sigma=4T_g^2\int d^2zD\left(z^2\right)$$
is the string tension of the Nambu-Goto term, and

$$\frac{1}{\alpha_0}=\frac{1}{4}T_g^4\int d^2zz^2\left(2D_1\left(z^2\right)
-D\left(z^2\right)\right)$$
is the inverse bare coupling constant of the rigidity term. Here
$\partial_a\equiv\frac{\partial}{\partial\xi^a}; a,b=1,2;
g_{ab}=\left(\partial_aw_\mu\right)\left(\partial_bw_\mu\right)$ 
is the induced metric tensor,
$g=det\parallel g_{ab} \parallel, t_{\mu\nu}=\frac{1}{\sqrt{g}}\varepsilon^
{ab}\left(\partial_aw_\mu\right)\left(\partial_bw_\nu\right)$ 
is the extrinsic curvature of
the string world sheet.

The aim of this letter is to apply action (1) to the derivation 
of the correction
to the Hamiltonian of the quark-antiquark system in the confining QCD
vacuum, which was obtained in$^{2}$ for the case of equal masses of a quark 
and antiquark 
and generalized in$^{3}$ to the case of arbitrary masses. Both in Refs.2  
and 3 only the Nambu-Goto term on the R.H.S. of Eq.(1) was
accounted for in the expression for the Green function of the spinless
$q\bar q$-system, written by virtue of the Feynman-Schwinger 
representation in the
form

$$G\left(x\bar x|y\bar y\right)
=\int\limits_0^\infty ds\int\limits_0^\infty d\bar s
\int Dz D\bar z~ e^{-K-\bar K} \left<W(C)\right>, \eqno (2)$$
where $K=m_1^2s+\frac{1}{4}\int\limits_0^s d\gamma\dot z^2(\gamma), 
\bar K=m_2^2\bar s+\frac{1}{4}\int\limits_0^{\bar s} d\gamma \dot{\bar
z}^2(\gamma)$, and our goal here is to consider also the rigidity term.
Analogously to Ref.3 we shall consider the $q\bar q$-system with
arbitrary masses of a quark and antiquark.

In this way we shall work within the same approximations, which were used
in$^{2,3}$, namely we shall disregard spin effects and the influence of
additional quark loops. Secondly, we shall neglect quark trajectories  
with backward motion in the proper time, which 
might lead to creation of additional $q\bar q$-pairs.  

 Besides that, we shall use the straight-line
approximation for the minimal surface $S$, that, as was argued in$^{2,3}$,
corresponds to the valence quark approximation. Such a ``minimal'' string
may rotate and oscillate longitudinally. This approximation is inspired by
two limiting cases: $l=0$ and $l\to\infty$. 

The first case
will be then investigated in more details, and 
the correction to the Hamiltonian of the relativistic quark model$^{9}$
due to the rigidity term in the limit of large masses of a quark 
and antiquark will be derived.

The main results of the letter are summarized in the Conclusion.

\vspace {6mm}
{\large \bf 2. A correction to the 
Hamiltonian of the ``minimal'' QCD string
with spinless quarks due to the rigidity term}

\vspace{3mm}
Making use of the auxiliary field formalism$^{2,10}$, one can represent 
Green function (2) with $\left<W(C)\right>$ 
defined via Eq.(1) in the following
way

$$G\left(x\bar x|y\bar y\right)=\int D\vec z D\vec{\bar z} D\mu_1D\mu_2Dh_{ab}
\exp\left(-K^\prime-\bar K^\prime\right)\exp\biggl[\left(-\sigma+2\bar\alpha
\right)\int
 d^2\xi\sqrt{h}{\,}\biggr]
\cdot$$

$$\cdot \exp\Biggl[-\bar\alpha\int d^2\xi \sqrt{h} h^{ab} \left(\partial_a 
w_\mu\right)
\left(\partial_b 
w_\mu\right)-\frac{1}{\alpha_0}\int d^2\xi \sqrt{h}h^{ab}\left(
\partial_a t_{\mu\nu}\right)
\left(\partial_b t_{\mu\nu}\right)\Biggr], \eqno(3)$$
where we have integrated over the Lagrange multiplier $\lambda^{ab}(\xi)=
\alpha(\xi)h^{ab}(\xi)+f^{ab}(\xi), f^{ab}h_{ab}=0$, and $\bar \alpha$
is the mean value of $\alpha(\xi)$.
Here $t_{\mu\nu}=\frac{1}{\sqrt{h}}\varepsilon^{ab}\left(\partial_a
w_\mu\right)\left(\partial_b w_\nu\right)$,

$$ K^\prime+\bar K^\prime=\frac{1}{2}\int\limits_0^T
d\tau\Biggl[\frac{m_1^2}{\mu_1(\tau)}+\mu_1(\tau)\left(1+
\dot{\vec z}{\,}^2(\tau)\right)+
\frac{m_2^2}{\mu_2(\tau)}+\mu_2(\tau)\left(1+\dot{\vec{\bar z}}{\,}^2(\tau)
\right)
\Biggr], \eqno (4)$$
$T=\frac{1}{2}
\left(x_0+\bar x_0-y_0-\bar y_0\right), 
\mu_1(\tau)=\frac{T}{2s}\dot z_0(\tau),
\mu_2(\tau)=\frac{T}{2\bar s}\dot{\bar z}_0(\tau)$, and the no-backtracking
time approximation$^{2,3}$

$$\mu_1(\tau)>0,~ \mu_2(\tau)>0 \eqno(5)$$
was used. Similarly to$^{2,3}$ we exploit in the valence quark sector (5) the
approximation that the minimal surface $S$ may be parametrized by the
straight lines, connecting points $z_\mu(\tau)$ and $\bar z_\mu(\tau)$ with
the same $\tau$, i.e. the trajectories of a quark and antiquark are
synchronized: $z_\mu=(\tau,\vec z), \bar z_\mu=(\tau, \vec{\bar z}),
w_\mu(\tau,\beta)=\beta z_\mu(\tau)+(1-\beta)\bar z_\mu(\tau), 0\le\beta
\le 1$.

Introducing auxiliary fields$^{2}$ $\nu(\tau,\beta)=T\sigma\frac{h_{22}}
{\sqrt{h}}, \eta(\tau,\beta)=\frac{1}{T}\frac{h_{12}}{h_{22}}$ and making
a rescaling $z_\mu\to\sqrt{\frac{\sigma}{2\bar\alpha}}z_\mu, \bar z_\mu\to
\sqrt{\frac{\sigma}{2\bar\alpha}}\bar z_\mu$ one gets from the last
exponent on the R.H.S. of Eq.(3) the following action of the string 
without quarks

$$A_{str.}=\int\limits_0^T d\tau\int\limits_0^1 d\beta\frac{\nu}{2}
\Biggl\{\dot w^2+\biggl(\biggl(\frac{\sigma}{\nu}\biggr)^2+\eta^2\biggr)
r^2-2\eta(\dot w r)+\frac{\sigma T^2}{\alpha_0\bar\alpha^2}
\frac{1}{h}\Biggl[\ddot w^2r^2-(\ddot w r)^2+\dot w^2\dot r^2-
(\dot w\dot r)^2+$$

$$+2((\ddot w\dot w)(\dot r r)-(\ddot w \dot r)(\dot w r))+
\biggl(\biggl(\frac{\sigma}{\nu}\biggr)^2+\eta^2\biggr)\left(\dot r^2 r^2-
(\dot r r)^2\right)-$$

$$-2\eta\left(
(\ddot w \dot r)r^2-(\ddot w r)(\dot r r)+(\dot w
\dot r)(\dot r r)-(\dot w r)\dot r^2\right)\Biggr]\Biggr\}, \eqno(6)$$
where a dot stands for $\frac{\partial}{\partial\tau}, r_\mu(\tau)=
z_\mu(\tau)-\bar z_\mu(\tau)$ is the relative coordinate, and analogously
to Ref.1 we have assumed that the string world sheet is not much
crumpled, so that $h_{ab}$ is a smooth function.

Let us now introduce the centre of masses coordinate $R_\mu(\tau)=
\zeta(\tau)z_\mu(\tau)+(1-\zeta(\tau))\bar z_\mu(\tau)$, where
$\zeta(\tau)\equiv\zeta_1(\tau)+\frac{1}{\alpha_0}\zeta_2(\tau), 
0\le \zeta(\tau)\le 1$, should be determined from the requirement 
that $\dot R_\mu$ decouples from $\dot r_\mu$ $^{3}$.
Next, assuming that a meson as a whole moves with a constant speed
(which is true for a free meson), i.e. $\ddot R=0$, and bringing
together quark kinetic terms (4) and pure string action (6),
we arrive at the following action of the QCD string with quarks 
 
$$A=\int\limits_0^T d\tau\Biggl\{\frac{m_1^2}{2\mu_1}+\frac{m_2^2}{2\mu_2}+
\frac{\mu_1}{2}+\frac{\mu_2}{2}+
\frac{1}{2}\left(\mu_1+\mu_2+\int\limits_0^1 d\beta \nu\right)\dot R^2+$$

$$+\left(\mu_1(1-\zeta_1)-\mu_2\zeta_1+\int\limits_0^1 d\beta (\beta-\zeta_1)
\nu\right)\left(\dot R\dot r\right)
-\int\limits_0^1 d\beta\nu\eta \left(\dot R r\right)+
\int\limits_0^1 d\beta (\zeta_1-\beta)\eta\nu (\dot r r)+$$

$$+\frac{1}{2}
\left(\mu_1(1-\zeta_1)^2+\mu_2\zeta_1^2+\int\limits_0^1 d\beta (\beta-
\zeta_1)^2\nu\right)\dot r^2+\frac{1}{2}\int\limits_0^1 d\beta\left(
\frac{\sigma^2}{\nu}+\eta^2\nu\right)r^2+$$

$$+\frac{1}{\alpha_0}
\Biggl[\zeta_2(\mu_1(\zeta_1-1)+\mu_2\zeta_1)\dot r^2-\zeta_2(\mu_1+\mu_2)
\left(\dot R \dot r\right)
+\int\limits_0^1 d\beta\nu\Biggl(\zeta_2(\zeta_1-\beta)
\dot r^2-\zeta_2\left(\dot R \dot r\right)
+\zeta_2\eta(\dot r r)+$$

$$+\frac{1}{2}(\beta-
\zeta_1)^2\left[\ddot{\vec r}, \vec r{\,}\right]^2
+\frac{1}{2}\dot R^2\dot r^2-\frac{1}{2}
\left(\dot R\dot r\right)^2+(\beta-\zeta_1)\left(\left(\ddot r \dot R\right)
(\dot r r)-(\ddot r\dot r)
\left(\dot R r\right)\right)
+\frac{1}{2}\biggl(\biggl(\frac{\sigma}{\nu}\biggr)^2+\eta^2
\biggr)\left[\dot{\vec r}, \vec r{\,}\right]^2+$$

$$+\eta\biggl((\beta-\zeta_1)\left((\ddot r r)
(\dot r r)-(\ddot r \dot r) r^2\right)
+\left(\dot R r\right)\dot r^2-\left(\dot R \dot r\right)(\dot r
 r)\biggr)\Biggr)\Biggr]\Biggr\}, \eqno (7)$$
where we have performed a rescaling $z_\mu\to\bar \alpha \sqrt{\frac{h}
{\sigma T^2}}z_\mu, \bar z_\mu\to\bar\alpha\sqrt{\frac{h}{\sigma T^2}}
\bar z_\mu, \nu\to\frac{\sigma T^2}{\bar\alpha^2 h}\nu$.

Integrating over $\eta$, one gets in the zeroth order in $\frac{1}{\alpha_0}~
\eta_{ext.}=\frac{(\dot r r)}{r^2}\biggl(\beta-\frac{\mu_1}{\mu_1+\mu_2}
\biggr)$, which together with the condition $\dot R\dot r=0$ yields
$\zeta_2^{ext.}=\frac{(\dot r r)^2}{r^2}\frac{\frac{\mu_1}{\mu_1+\mu_2}
\int\limits_0^1 d\beta\nu-\int\limits_0^1 d\beta\beta\nu}{\mu_1+\mu_2+
\int\limits_0^1 d\beta\nu}$, while $\zeta_1^{ext.}=\frac{\mu_1+\int
\limits_0^1 d\beta\beta\nu}{\mu_1+\mu_2+\int\limits_0^1 d\beta\nu}$ was
found in$^{3}$.

Finally, in order to obtain the desirable Hamiltonian, 
we shall perform the usual
canonical transformation from $\dot{\vec R}$ to the total momentum
$\vec P$ in the Minkowski space-time

$$\int D\vec R \exp\biggl[i\int L\left(\dot{\vec R},...\right)d\tau\biggr]=
\int D\vec R D\vec P \exp\biggl[i\int\left(\vec P\dot{\vec R}- H\left(
\vec P,...\right)\right)
d\tau\biggr],$$
where $H\left(\vec P,...\right)=\vec P\dot{\vec R}- L\left(\dot{\vec R},...
\right)$, and
choose the meson rest frame as $\vec P=\frac{\partial L\left(\dot{\vec R}
,...\right)}{\partial \dot{\vec R}}=0$. After performing the transformation
from $\dot{\vec r}$ to $\vec p$ we get the following Hamiltonian

$$H=H^{(0)}+\frac{1}{\alpha_0}H^{(1)}. \eqno (8)$$
Here
$$H^{(0)}=\frac{1}{2}\left[\frac{\left(\vec p_r{\,}^2+m_1^2\right)}{\mu_1}+
\frac{\left(\vec p_r{\,}^2+m_2^2\right)}{\mu_2}+\mu_1+\mu_2+\sigma^2
\vec r{\,}^2\int\limits_0^1\frac{d\beta}{\nu}+\nu_0+
\frac{\vec L{\,}^2}{\rho \vec r{\,}^2}\right] \eqno (9)$$
with

$$\rho= \mu_1+\nu_2-\frac{(\mu_1+\nu_1)^2}{\mu_1+\mu_2+\nu_0},~ 
\nu_i\equiv\int\limits_0^1 d\beta\beta^i \nu,~ \vec p_r{\,}^2\equiv
\frac{(\vec p{\,}\vec r{\,})^2}{\vec r{\,}^2},~ \vec L\equiv \left[\vec r, 
\vec p{\,}\right]$$
is the Hamiltonian of the ``minimal'' Nambu-Goto string with quarks,
which was derived and investigated in$^{3}$, while the new Hamiltonian
$H^{(1)}$ has the form

$$H^{(1)}=\frac{a_1}{\rho^2}\vec L{\,}^2+
\frac{a_2}{\rho^2}\dot{\vec L}{\,}^2+
\frac{a_3}{2\tilde\mu^3}\left | \vec r{\,}\right |\left(\vec p_r{\,}^2\right)
^{\frac{3}{2}}+
\frac{a_4}{\tilde \mu^4}\left(\vec p_r{\,}^2\right)^2+
\frac{a_5}{2\tilde\mu\rho^2}\frac{\sqrt{\vec p_r{\,}^2}\vec L{\,}^2}
{\left| \vec r{\,} 
\right|}+
\frac{a_6}{2\tilde \mu^2 \rho^2}\frac{\vec p_r{\,}^2 \vec L{\,}^2}
{\vec r{\,}^2}, \eqno (10)$$
where 
$\tilde \mu=\frac{\mu_1\mu_2}{\mu_1+\mu_2}$, and the
coefficients $a_k, k=1,...,6$ read as follows

$$a_1=\frac{\sigma^2}{2}\int\limits_0^1\frac{d\beta}{\nu},~ a_2=\frac{1}{2}
\Biggl[\nu_2+\frac{(\mu_1+\nu_1)(\nu_0(\mu_1-\nu_1)-2\nu_1(\mu_1+
\mu_2))}{(\mu_1+\mu_2+\nu_0)^2}\Biggr],~ 
a_3=3\frac{\dot{\tilde\mu}}{\tilde\mu}B-\dot B,$$

$$a_4=\frac{1}{2(\mu_1+\mu_2)(\mu_1+\mu_2+\nu_0)}
\Biggl[\frac{\nu_0(\mu_1\nu_0-\nu_1(\mu_1+\mu_2))^2}{(\mu_1+\mu_2)
(\mu_1+\mu_2+\nu_0)}-\nu_1(\mu_1+\mu_2)(\nu_1-2\mu_2)-\mu_1\nu_0
(\mu_1+2\mu_2)\Biggr],$$

$$a_5=\frac{\dot{\tilde\mu}}{\tilde\mu}B-\dot B,$$

$$a_6=\nu_2+\frac{\nu_1^2+2\mu_1\nu_0-2\mu_2\nu_1}{\mu_1+
\mu_2+\nu_0}+\frac{1}{\mu_1+\mu_2}\Biggl[\frac{1}{\mu_1+\mu_2}
\Biggl(\frac{(\mu_1\nu_0-\nu_1(\mu_1+\mu_2))^2(3\nu_0+2(\mu_1+\mu_2))}
{(\mu_1+\mu_2+\nu_0)^2}+$$

$$+\mu_1(\mu_1\nu_0-2\nu_1(\mu_1+\mu_2))\Biggr)-
\frac{\mu_1^2\nu_0}{\mu_1+\mu_2+\nu_0}\Biggr].$$
Here $B\equiv\frac{\nu_1(\mu_1+\mu_2)(\nu_1-2\mu_2)+\mu_1\nu_0(\mu_1+
2\mu_2)}{(\mu_1+\mu_2)(\mu_1+\mu_2+\nu_0)}$, and 
during the derivation of $H^{(1)}$ we have chosen the origin 
at the centre of masses of the initial state,
so that $\dot{\vec R}\dot{\vec r}\ll 1$, and the term 
$-\frac{1}{2\alpha_0}\int\limits_0^T d\tau\nu_0\left(\dot{\vec R}\dot
{\vec r}{\,}\right)^2$
on the R.H.S. of Eq.(7) has been neglected. 

Notice, that Hamiltonian (8)-(10) contains auxiliary fields $\mu_1$,
$\mu_2$ and $\nu$. In order to construct the operator Hamiltonian, 
acting onto wave functions, one should integrate over these fields (that 
implies the substitution of their extremal values, which could be obtained 
from the corresponding saddle point equations, into Eqs.(8)-(10)) and 
perform Weil ordering$^{11}$.

Let us now 
apply Hamiltonian (10) to the derivation of the rigid string
correction to the Hamiltonian of the so-called 
relativistic quark model$^{9}$, i.e. consider the case when the orbital 
momentum is equal to zero. In what follows we shall 
put for simplicity the mass of a quark being equal to the mass 
of an antiquark $m_1=m_2\equiv m$. 
In order to get $H^{(1)}$, one should substitute the extremal values of 
the fields $\mu_1$, $\mu_2$ and $\nu$  
of the zeroth order in $\frac{1}{\alpha_0}~
\mu_1^{ext.}=\mu_2^{ext.}=\sqrt{\vec p{\,}^2+m^2}$ and $\nu_{ext.}=
\sigma\left|
\vec r{\,}\right|$ 
into Eq.(10). 
The limit of large masses of a quark and antiquark means that $m\gg\sqrt{
\sigma}$. In this case 
we obtain from Eq.(10) 
the rigid string Hamiltonian $H^{(1)}=-\frac{4\sigma\left|\vec r{\,}\right|}
{m^4}\left(\vec p_r
{\,}^2\right)^2$ and then from Eqs.(8) and (9) the following expression 
for the 
total Hamiltonian $H$ 

$$H=2m+\sigma\left|\vec r{\,}\right|+\frac{\vec p{\,}^2}{m}-
\left(\frac{1}{4m^3}+\frac{4\sigma\left|\vec r{\,}\right|}
{\alpha_0 m^4}\right)\left(\vec p{\,}^2\right)^2. \eqno (11)$$

\vspace{6mm}
{\large \bf 4. Conclusion}

\vspace{3mm}
In this letter we have derived a correction to Hamiltonian (9) of the
QCD string with spinless quarks, which was found in Ref.3, arising
due to the rigidity term in the gluodynamics string effective action$^{1}$.
This correction is given by formula (10). 
The Hamiltonian obtained contains corrections to the 
orbital momentum of the system and also several operators higher than 
of the second order in the momentum. The latter ones arise due to the 
fact that the rigid string theory is a theory with higher derivatives.

Making use of the obtained Hamiltonian we have derived a rigid string 
contribution 
to
the Hamiltonian of the relativistic quark model in the case of equal 
large masses of a quark and antiquark, so that the total Hamiltonian 
is given by formula 
(11).

\vspace{6mm}
{\large \bf 5. Acknowledgments}

\vspace{3mm}
I am grateful to Professors D.Ebert, M.G.Schmidt and Yu.A.Simonov for 
useful discussions and to Professor Yu.M.Makeenko for his permanent 
interest to my work. 
I would also like to thank 
the theory group of the Quantum Field Theory Department of the 
Institut f\"ur Physik of the Humboldt-Universit\"at of Berlin for kind 
hospitality.

\newpage
{\large\bf References}

\vspace{3mm}
\noindent
1.~ D.V.Antonov, D.Ebert and Yu.A.Simonov, {\it Mod.Phys.Lett.} {\bf A11}, 
1905 (1996).\\
2.~ A.Yu.Dubin, A.B.Kaidalov and Yu.A.Simonov, {\it Yad.Fiz.} {\bf 56}, 
213 (1993), 
{\it Phys.Lett.} {\bf B323}, 
41 (1994).\\
3.~ E.L.Gubankova and A.Yu.Dubin, {\it Phys.Lett.} {\bf B334}, 180 (1994), 
preprint ITEP 62-94.\\
4.~ M.B.Halpern, {\it Phys.Rev.} {\bf D19}, 517 (1979); I.Ya.Aref'eva, 
{\it Theor.Math.Phys.} {\bf 43}, 111 (1980); N.Bralic, {\it Phys.Rev.} 
{\bf D22}, 3090 (1980).\\
5.~ Yu.A.Simonov, {\it Yad.Fiz.} {\bf 50}, 213 (1989).\\
6.~ N.G. Van Kampen, {\it Stochastic Processes in Physics and Chemistry} 
(North-Holland, Amsterdam, 1984).\\
7.~ Yu.A.Simonov, {\it Yad.Fiz.} {\bf 54}, 192 (1991).\\
8.~ H.G.Dosch, {\it Phys.Lett.} {\bf B190}, 177 (1987); Yu.A.Simonov, 
{\it Nucl.Phys.} {\bf B307}, 512 (1988); H.G.Dosch and Yu.A.Simonov, 
{\it Phys.Lett.} {\bf B205}, 339 (1988), {\it Z.Phys.} {\bf C45}, 147 
(1989); Yu.A.Simonov, {\it Nucl.Phys.} {\bf B324}, 67 (1989), 
{\it Phys.Lett.} {\bf B226}, 151 (1989), {\it Phys.Lett.} {\bf B228}, 413 
(1989).\\
9.~ P.Cea, G.Nardulli and G.Preparata,
{\it Z.Phys.} {\bf C16}, 135 (1982), {\it Phys.Lett.} 
{\bf B115}, 310 (1982); J.Carlson et al., {\it Phys.Rev.} {\bf D27}, 233 
(1983); J.L.Basdevant and S.Boukraa, {\it Z.Phys.} {\bf C30}, 103 (1986); 
S.Godfrey and N.Isgur, {\it Phys.Rev.} {\bf D32}, 189 (1985).\\  
10. A.M.Polyakov, {\it Gauge Fields and Strings} (Harwood, 1987).\\
11. T.D.Lee, {\it Particle Physics and Introduction to Field Theory} 
(Harwood, 1990).
\end{document}